\documentclass[a4paper, UKenglish, cleveref, autoref, thm-restate]{lipics-v2021}
\hideLIPIcs

\usepackage{algorithm}
\usepackage[noend]{algpseudocode}

\algnewcommand\algorithmicforeach{\textbf{foreach}}
\algdef{S}[FOR]{ForEach}[1]{\algorithmicforeach\ #1\ \algorithmicdo}

\title{Reducing Profile-Based Matching to the Maximum Weight Matching Problem}

\author{Seongbeom Park}{Ulsan National Institute of Science and Technology (UNIST), South Korea}{sparkamita90@gmail.com}{https://orcid.org/0000-0002-7012-6587}{}

\authorrunning{S. Park}

\Copyright{Seongbeom Park}



\category{}

\relatedversion{}

\nolinenumbers

\begin{document}

\maketitle

\begin{abstract}
The profile-based matching problem is the problem of finding a matching that optimizes profile from an instance $(G, r, \langle u_1, \dots, u_r \rangle)$, where $G$ is a bipartite graph $(A \cup B, E)$, $r$ is the number of utility functions, and $u_i: E \to \{ 0, 1, \dots, U_i \}$ is utility functions for $1 \le i \le r$. A matching is optimal if the matching maximizes the sum of the 1st utility, subject to this, maximizes the sum of the 2nd utility, and so on. The profile-based matching can express rank-maximal matching \cite{irving2006rank}, fair matching \cite{huang2016fair}, and weight-maximal matching \cite{huang2012weight}. These problems can be reduced to maximum weight matching problems, but the reduction is known to be inefficient due to the huge weights.

This paper presents the condition for a weight function to find an optimal matching by reducing profile-based matching to the maximum weight matching problem. It is shown that a weight function which represents utilities as a mixed-radix numeric system with base-$(2U_i+1)$ can be used, so the complexity of the problem is $O(m\sqrt{n}(\log{n} + \sum_{i=1}^{r}\log{U_i}))$ for $n = |V|$, $m = |E|$. In addition, it is demonstrated that the weight lower bound for rank-maximal/fair/weight-maximal matching, better computational complexity for fair/weight-maximal matching, and an algorithm to verify a maximum weight matching can be reduced to rank-maximal matching. Finally, the effectiveness of the profile-based algorithm is evaluated with real data for school choice lottery.
\end{abstract}

\section{Introduction}
The profile-based matching problem is given as input an instance $(G, r, \langle u_1, \dots, u_r \rangle)$ with a bipartite graph $G = (A \cup B, E)$, the number of utility functions $r$, and a non-negative utility functions $u_i: E \to \{ 0, 1, \dots, U_i \}$ for $1 \le i \le r$.
Each utility function is a function that quantifies some feature such as preference, commuting distance, or network latency, according to the problem domain.
The utility functions have priorities, with $u_1$ having the highest priority, followed by $u_2$, and so on.
A matching is optimal if the matching maximizes the sum of the 1st utility, subject to this, maximizes the sum of the 2nd utility, and so on.


The profile-based matching can express rank-maximal matching \cite{irving2006rank}, fair matching \cite{huang2016fair}, and weight-maximal matching \cite{huang2012weight} by utility function definition.
Although these problems can be reduced to the maximum weight matching problem \cite{irving2006rank,huang2012weight,manlove2013algorithmics,huang2016fair}, reduction is expensive since the weights are steeply increasing.
To reduce the reduction complexity for rank-maximal matching, scaling algorithm \cite{michail2007reducing} and the lower bound of the weights \cite{berczi2022approximation} have been studied.

This paper presents the condition for a weight function to find an optimal matching by reducing profile-based matching to the maximum weight matching problem.
It is shown that a weight function which represents utilities as a mixed-radix numeric system with base-$(2U_i+1)$ can be used, so the complexity of the problem is $O(m\sqrt{n}(\log{n} + \sum_{i=1}^{r}\log{U_i}))$ for $n = |V|$, $m = |E|$.
In addition, it is demonstrated that the weight lower bound for rank-maximal/fair/weight-maximal matching, better computational complexity for fair/weight-maximal matching, and an algorithm to verify a maximum weight matching can be reduced to rank-maximal matching.
Finally, the effectiveness of the profile-based algorithm is evaluated with real data for school choice lottery.

\begin{description}
\item[Contribution 1] Proving the condition of weight function to reduce profile-based matching problem to the maximum weight matching problem

This paper proves that the maximum weight matching is the profile optimal matching when the sum of weights is increasing as the profile of matching is increases.
The proof is simplified by converting an arbitrary bipartite graph to a complete bipartite graph.
The condition suggests weight lower bound for the related problems; such as the weight ratio of rank-maximal matching \cite{berczi2022approximation} can be less than 2, fair matching \cite{huang2016fair} and weight-maximal matching \cite{huang2012weight} can be found with a weight function $w(e) = \sum_{i=1}^{r}{u_i(e) \prod_{j=i+1}^{r}{(2U_j+1)}}$ which is independent to $n$.

\item[Contribution 2] Improving the computational complexity for related problems

This paper presents an algorithm which achieves better complexity for related problems by reduction with lesser weights.
While the complexity of both fair matching and weight-maximal matching are known to be $O(r^*m\sqrt{n}\log{n})$ when $r^*$ is the maximum rank of the matching \cite{huang2016fair,huang2012weight,manlove2013algorithmics}, the complexity of algorithm proposed in this paper is $O(m\sqrt{n}(\log{n} + r))$.
This paper also presents another algorithm to verify a maximum weight matching can be reduced rank-maximal matching problem.
If reduction is available, the maximum weight matching problem can be solved in $O(r^*m\sqrt{n})$.

\item[Contribution 3] Evaluating profile-based matchings with real data

Rank-maximal matching (RM) is known to be an effective method to increase preference satisfaction in school choice lottery \cite{featherstone2020rank,ortega2023cost}.
This paper quantitatively evaluates RM and minimum-cost rank-maximal matching (MCRM), which minimizes the commuting distance, with real students' data in $\alpha$-city.
Experimental results show that RM and MCRM assign all students to their first or second choice.
MCRM achieves the lowest total commuting distance, 4.18\% and 6.15\% less than that of Baseline and RM, respectively.
\end{description}

In the following \cref{sec:preliminaries}, the profile-based matching problem is defined.
In \cref{sec:reduction}, the condition of weight function is proved.
\cref{sec:complexity} presents the reduction algorithm and its complexity.
\cref{sec:related_problems} deals the lower bound and complexity for related problems.
\cref{sec:experiment} shows the experimental results using real data.

\section{Preliminaries}\label{sec:preliminaries}
\emph{The profile-based matching problem} is the problem of finding a matching that optimizes the profile from an instance $(G, r, \langle u_1, \dots, u_r \rangle)$, where a bipartite graph $G = (A \cup B, E)$, $r$ number of utility functions, and utility function $u_i$.
The \emph{utility function} is defined as $u_i: E \to \{ 0, 1, \dots, U_i \}$, for $1 \le i \le r$.
A \emph{matching} is a subset of $E$, where no edges are incident to each other.
If a pair $(a, b)$ is in matching $M$, then $a$ and $b$ are \emph{partners} to each other, and denoted as $M(a) = b$, $M(b) = a$.
The \emph{profile} of a matching $p(M)$ is defined as a $r$-tuple $\langle p_1(M), \dots, p_r(M) \rangle$, which $p_i(M) = \sum_{e \in M}{u_i(e)}$ for $1 \le i \le r$.
The profiles are sorted lexicographically, and are denoted by $p(M) > p(M')$ for matchings $M, M'$ if there exists $j \le r$ such that $p_j(M) > p_j(M')$, where $p_i(M) = p_i(M')$ for all $i < j$.
If $p_i(M) = p_i(M')$ for $1 \le i \le r$, then $p(M) = p(M')$.
\emph{The optimal matching} is defined as a matching which profile is greater than or equal to that of all other possible matchings.

\begin{lemma}\label{lemma:perfect}
The intersection of the optimal matching for $(G', r, \langle u'_1, \dots, u'_r \rangle)$ and $E$ is the optimal matching from $(G, r, \langle u_1, \dots, u_r \rangle)$, where $G'$ is the complete bipartite graph $(A \cup B, A \times B)$, $u_i'(e)$ is 0 for $e \notin E$ otherwise $u_i(e)$, and $G$ is the original graph $(A \cup B, E)$.
\end{lemma}

\begin{proof}
Let matching $M'$ is optimal matching for $(G', r, \langle u'_1, \dots, u'_r \rangle)$, and matching $M = M' \cap E$.

Let assume that matching $M$ is not optimal.
There exist a matching $M^*$ which satisfies $p(M^*) > p(M)$.
Since $\forall e \notin E,\ u_i'(e) = 0$, matchings $M, M', M^*$ satisfy the following:
$$p(M') = p(M' \cap E) = p(M) < p(M^*)$$

Since $M^* \subset E \subset A \times B$, matching $M^*$ is also a matching of $(G', r, \langle u'_1, \dots, u'_r \rangle)$ which profile is greater than matching $M'$.
The assumption that matching $M$ is not an optimal matching creates a contradiction that matching $M'$ is not optimal matching.
Therefore, the intersection of the optimal matching of the complete bipartite graph and the set of edges of the original graph is optimal matching.
\end{proof}

By \cref{lemma:perfect}, the problem of finding profile-based optimal matching in an arbitrary bipartite graph becomes the same as the problem of finding optimal matching from the complete bipartite graph, and utility functions which is 0 for additional pairs.
From now on, a complete bipartite graph is assumed unless explicitly stated for simplicity.

A pair $(a, b) \notin M$ is \emph{improving pair} for matching $M$ when the pair satisfying $p(\{ (a, b) \}) > p(\{ (a, M(a)), (M(b), b) \})$.
Let the operation is called \emph{improve} when creating a new matching by replacing the partners of the improving pair.

\begin{lemma}
Improved matching has greater profile than the original matching.
\end{lemma}

\begin{proof}
Let pair $(a, b)$ is an improving pair for matching $M$.
Matching $M$ and improved matching $M'$ satisfies the following equations:
$$M \setminus \{(a, M(a)), (M(b), b)\} = M' \setminus \{(a, b), (M(b), M(a))\}$$
$$p(M) - p(\{(a, M(a)), (M(b), b)\}) = p(M') - p(\{(a, b), (M(b), M(a))\})$$

Since the utility function is a non-negative function, the following inequalities are satisfied by the definition of improving pair:
$$p(\{(a, M(a)), (M(b), b) \}) < p(\{(a, b)\}) \le p(\{(a, b), (M(b), M(a))\})$$
$$\therefore p(M) < p(M')$$
\end{proof}

\section{Reduction}\label{sec:reduction}
In this section, the condition of weight function is presented to reduce a profile-based matching to the maximum weight matching.
If the weight function is designed so that the sum of the weights of the matching increases as the profile of the matching increases, the maximum weight matching becomes optimal matching, which is the matching with the greatest profile.
Since the profile of matching increases by improving, the weight function should return greater weight to the potentially improving pair.

\begin{theorem}\label{theorem:weight}
The maximum weight matching is optimal matching, if the weight function $w$ satisfies $p(\{ (a, b) \}) > p(\{ (a, b'), (a', b) \}) \implies w((a, b)) > w((a, b')) + w((a', b)) \ge 0$.
\end{theorem}

\begin{proof}
Let assume that matching $M$ is not optimal matching though it is the maximum weight matching generated by the weight function $w$.
There is an improving pair $(a, b)$ for matching $M$, and let matching $M'$ be the improved matching.
Matchings $M, M'$ satisfy the following equations:
$$M \setminus \{(a, M(a)), (M(b), b)\} = M' \setminus \{(a, b), (M(b), M(a))\}$$
$$\sum\limits_{e \in M}{w(e)} - (w((a, M(a))) + w((M(b), b))) = \sum\limits_{e \in M'}{w(e)} - (w((a, b)) + w((M(b), M(a))))$$

Since pair $(a, b)$ is improving pair for matching $M$ and weight function $w$ is non-negative function, the following inequalities are satisfied:
$$p(\{(a, M(a)), (M(b), b) \}) < p(\{(a, b)\})$$
$$w((a, M(a))) + w((M(b), b)) < w((a, b)) \le w((a, b)) + w((M(b), M(a)))$$
$$\therefore \sum\limits_{e \in M}{w(e)} < \sum\limits_{e \in M'}{w(e)}$$

There is no matching with a sum of weights greater than the maximum weight matching.
However, the assumption that the maximum weight matching is not optimal matching creates a contradiction that there exists a matching $M'$ with a sum of weights greater than the maximum weight matching $M$.
Therefore, the maximum weight matching generated by the weight function $w$ is optimal matching.
\end{proof}

\begin{theorem}
The weight function $w_{mixed}$, which assigns the weight of an edge as a mixed-radix number of base-$(2U_i+1)$, can be used for finding an optimal matching.
\end{theorem}

\begin{proof}
The weight function $w_{mixed}$ can be expressed with utility functions as follow:
\begin{align*}
\qquad\qquad\qquad\qquad
w_{mixed}(e) &= u_1(e)_{(2U_1+1)} u_2(e)_{(2U_2+1)} \dots u_r(e)_{(2U_r+1)} \\
&= \sum\limits_{i=1}^{r}{u_i(e) \prod\limits_{j=i+1}^{r}{(2U_j+1)}}
\end{align*}

Let pair $(a, b)$ is a potentially improving pair.
There exists an integer $x \le r$ such that:
$$u_x((a, b)) > u_x((a, b')) + u_x((a', b))$$
$$u_i((a, b)) = u_i((a, b')) + u_i((a', b)) \text{ for all } i < x$$

The positions after $x$ of $w_{mixed}((a, b')) + w_{mixed}((a', b))$ is at most ${2U_{x+1}}_{(2U_{x+1}+1)} \dots {2U_{r}}_{(2U_{r}+1)}$, which is less than $1_{(2U_{x}+1)} 0_{(2U_{x+1}+1)} \dots 0_{(2U_{r}+1)}$.
Therefore, weight function $w_{mixed}$ satisfies $w_{mixed}((a, b)) > w_{mixed}((a, b')) + w_{mixed}((a', b)) \ge 0$, if $p(\{ (a, b) \}) > p(\{ (a, b'), (a', b) \})$.
\end{proof}

\section{Complexity Analysis}\label{sec:complexity}
\begin{algorithm}[ht]
    \caption{An algorithm for finding a profile-based optimal matching by reduction.}\label{alg:reduction}
    \begin{algorithmic}[1]
        \Function{getProfileBasedOptimalMatching}{$V, E, r, \langle u_1, \dots, u_r \rangle, \langle U_1, \dots U_r \rangle$}
            \State $w \gets new\ Map(key\_type = Edge, value\_type = Integer)$ \label{code:weight}
            \ForEach{$e \in E$}
                \State $w[e] \gets \sum\limits_{i=1}^{r}{u_i(e) \prod\limits_{j=i+1}^{r}{(2U_j+1)}}$
            \EndFor \label{code:weight_done}
            \State $optimal\_matching \gets getMaximumWeightMatching(V, E, w)$ \label{code:mwm}
            \State \Return $optimal\_matching$
        \EndFunction
    \end{algorithmic}
\end{algorithm}

\cref{alg:reduction} shows an algorithm that solves profile-based matching problem by reducing to the maximum weight matching.
Let $n = |V|$, $m = |E|$.
Line~\ref{code:weight}-\ref{code:weight_done} calculates the weights of edges with the utility functions.
The conversion uses weight function $w_{mixed}$, which considers the utilities of an edge as a mixed radix number with base-$(2U_i+1)$.
It takes $O(rm)$ to compute the weights of all edges.
Line~\ref{code:mwm} uses the converted weights to find a maximum weight matching.
The complexity of Line~\ref{code:mwm} is $O(m\sqrt{n}\log{(n N)})$, when $N$ is the maximum value of the weight \cite{gabow1989faster}.
Since the maximum weight is less than $\prod_{i=1}^{r}{(2U_i+1)}$, the complexity of \cref{alg:reduction} is $O(m\sqrt{n}(\log{n} + \sum_{i = 1}^{r}\log{U_i}))$.

\section{Related Problems}\label{sec:related_problems}
\subsection{Weight lower bound for rank-maximal matching}
It is known that the ratio of weights should be greater than 2 to reduce rank-maximal matching to maximum weight matching \cite{berczi2022approximation}.
However, there are corner cases that the ratio of weights is less than 2 while satisfying \cref{theorem:weight}.
Let assume a global preference system, which the rank of an edge is determined by a single preference list \cite{gai2007acyclic}, with constraint that all edges belonging to a certain rank are shares a common vertex.
If the weight function $w_{GRP}$ is defined as follows for edge $e_i$ which rank is $i$, $\frac{w_{GRP}(e_{i})}{w_{GRP}(e_{i+1})}$ is less than 2, for $1 \le i \le r - 3$ when $r \ge 4$.
\begin{align*}
\qquad\qquad\qquad\qquad\qquad
w_{GRP}(e_{i}) &= w_{GRP}(e_{i+1}) + w_{GRP}(e_{i+2}) + 1 \\
w_{GRP}(e_{r-1}) &= 2 \\
w_{GRP}(e_{r}) &= 1
\end{align*}

\subsection{Complexity of fair matching}
The fair matching problem \cite{huang2012weight,manlove2013algorithmics,huang2016fair} with maximum rank is $r$, can be represented with $r+1$ utility functions $u_{fair, i}$ for $1\le j \le r$:
$$u_{fair, 1}(e) = 1$$
$$
u_{fair, j+1}((a, b)) = \begin{cases}
2 & \text{if both $a$ and $b$ rank each other as rank $\le r - j + 1$}\\
1 & \text{if one of $a$ or $b$ ranks the other as rank $\le r - j + 1$} \\
0 & \text{otherwise}
\end{cases}
$$
\cref{alg:reduction} achieves the complexity of $O(m\sqrt{n}(\log{n} + r))$, since the maximum value of utility functions are 1 or 2, while the complexity of combinatorial algorithm in \cite{huang2016fair} is $O(r^*m\sqrt{n}\log{n})$ where $r^*$ is the maximum rank of edge in the matching.

\subsection{Complexity of weight-maximal matching}
The weight-maximal matching problem \cite{huang2012weight} can be represented by using utility functions with the same maximum utility $W$.
Since $W$ is assumed as a constant, \cref{alg:reduction} achieves the complexity of $O(m\sqrt{n}(\log{n} + r))$, while the complexity of combinatorial algorithm in \cite{huang2012weight} is $O(r^*m\sqrt{n}\log{n})$.

\subsection{Complexity of maximum weight matching}
The combinatorial algorithm for rank-maximal matching \cite{irving2006rank} has a lower complexity $O(r^*m\sqrt{n})$ than \cref{alg:reduction}, because it solves the problem by iterating maximum cardinality matching.
If an instance of maximum weight matching can be reduced to rank-maximal, it may save much of computation time.
\cref{alg:sort_sweep} is an algorithm that verifies whether the instance of the maximum weight matching can be reduced to a rank-maximal matching.

\begin{algorithm}[ht]
    \caption{An algorithm to verify the rank-maximal condition.}\label{alg:sort_sweep}
    \begin{algorithmic}[1]
        \Function{isRankMaximal}{$V, E, w$}
            \State $weight\_list \gets new\ List(type = Integer)$ \label{code:sort}
            \ForEach{$e \in E$}
                \State $weight\_list.append(w(e))$
            \EndFor
            \State $sorted\_list \gets sort(weight\_list, increasing = True)$ \label{code:sort_done}
            \State $\langle w_i, w_j, w_k \rangle \gets \langle sorted\_list[0], 0, 0 \rangle$ \label{code:sweep}
            \For{$i \gets 1 \text{ to } |E|$}
                \State $\langle w_i, w_j, w_k \rangle \gets \langle sorted\_list[i], w_i, w_j \rangle$
                \If{$w_i \le w_j + w_k$}
                    \State \Return $False$
                \EndIf
            \EndFor \label{code:sweep_done}
            \State \Return $True$
        \EndFunction
    \end{algorithmic}
\end{algorithm}

Line~\ref{code:sort}-\ref{code:sort_done} sort the weights of edges in $O(m \log{n})$, since $m \le (\frac{n}{2})^2$.
Line~\ref{code:sweep}-\ref{code:sweep_done} sweep the sorted weights and check whether they satisfy the condition of rank-maximal.
Therefore, the overall complexity of \cref{alg:sort_sweep} is $O(m \log{n})$.
If the algorithm returns $True$, then the problem can be reduced to a rank-maximal matching problem by setting the edges with the greatest weighted to rank $1$, the edges with the next greatest weighted to rank $2$, ..., the edges with the least weighted to rank $r$.

\section{Experimental Results}\label{sec:experiment}
In this section, the effectiveness of profile-based matching is evaluated with real data on students' preferences, commuting distances, and schools' quotas which was used for actual school choice lottery in $\alpha$-city in 2023.
There are 9 schools in the city, and each school has separate quotas for males and females.
Table~\ref{tab:quota} shows the number of male and female students in each school.

\begin{table}[ht] 
    \centering
    \begin{tabular}{|c|c|c|c|}
        \hline
        School  & Male  & Female    & Total \\
        \hline
        \hline
        $h_1$   & 154   & 0         & 154   \\
        $h_2$   & 161   & 0         & 161   \\
        $h_3$   & 172   & 0         & 172   \\
        $h_4$   & 0     & 186       & 186   \\
        $h_5$   & 0     & 170       & 170   \\
        $h_6$   & 0     & 172       & 172   \\
        $h_7$   & 80    & 73        & 153   \\
        $h_8$   & 77    & 46        & 123   \\
        $h_9$   & 71    & 52        & 123   \\
        \hline
        Total   & 715   & 699       & 1414  \\
        \hline
    \end{tabular}
    \caption{School quotas in $\alpha$-city in 2023.}
    \label{tab:quota}
\end{table}

Students applied three schools.
Students who were not assigned to a school to which they applied due to insufficient seats could be assigned to another school to which they did not apply.
In other words, schools not included in the choices are considered as fourth choices, which forms complete preferences with ties.


\subsection{Baseline: Actual assignment result in $\alpha$-city in 2023}
Baseline shows the actual result in $\alpha$-city in 2023.
The assignment was divided into two stages.
In the first stage, 80\% of the seats in each school were allocated based on the students' preferences using a greedy algorithm.
In the next stage, the schools were ranked in six ranks from the nearest school to the furthest school for each student, and the remaining 20\% of seats were assigned using a greedy algorithm using the rank determined by distance rank.

\subsection{RM: Rank-maximal matching}
The rank-maximal matching (RM) consider only the preferences of students, so commuting distance is not taken into account.
The weight function $w_{RM}(e) = 2^{4 - rank(e) + 1} - 1$ is used for RM which satisfies \cref{theorem:weight}.

\subsection{MCRM: Minimum-cost rank-maximal matching}
The minimum-cost rank-maximal matching (MCRM) is a matching which minimizes the total commuting distance among rank-maximal matchings.
Commuting distance is calculated by using the walking distance\footnote{The average of the recommended walking distance on Naver Map and the shortest walking distance on Kakao Map is measured in 0.01km units} from the student's residence to the school.
For the commuting distance function $d: E \to \{0, 1, \dots, D\}$, the weight function $w_{MCRM}(e) = (D+1)2^{4 - rank(e)} - D - d(e)$ is used for experiment.

\subsection{Profile comparison}
\begin{figure}[ht]
    \centering
    \includegraphics[width=0.95\linewidth]{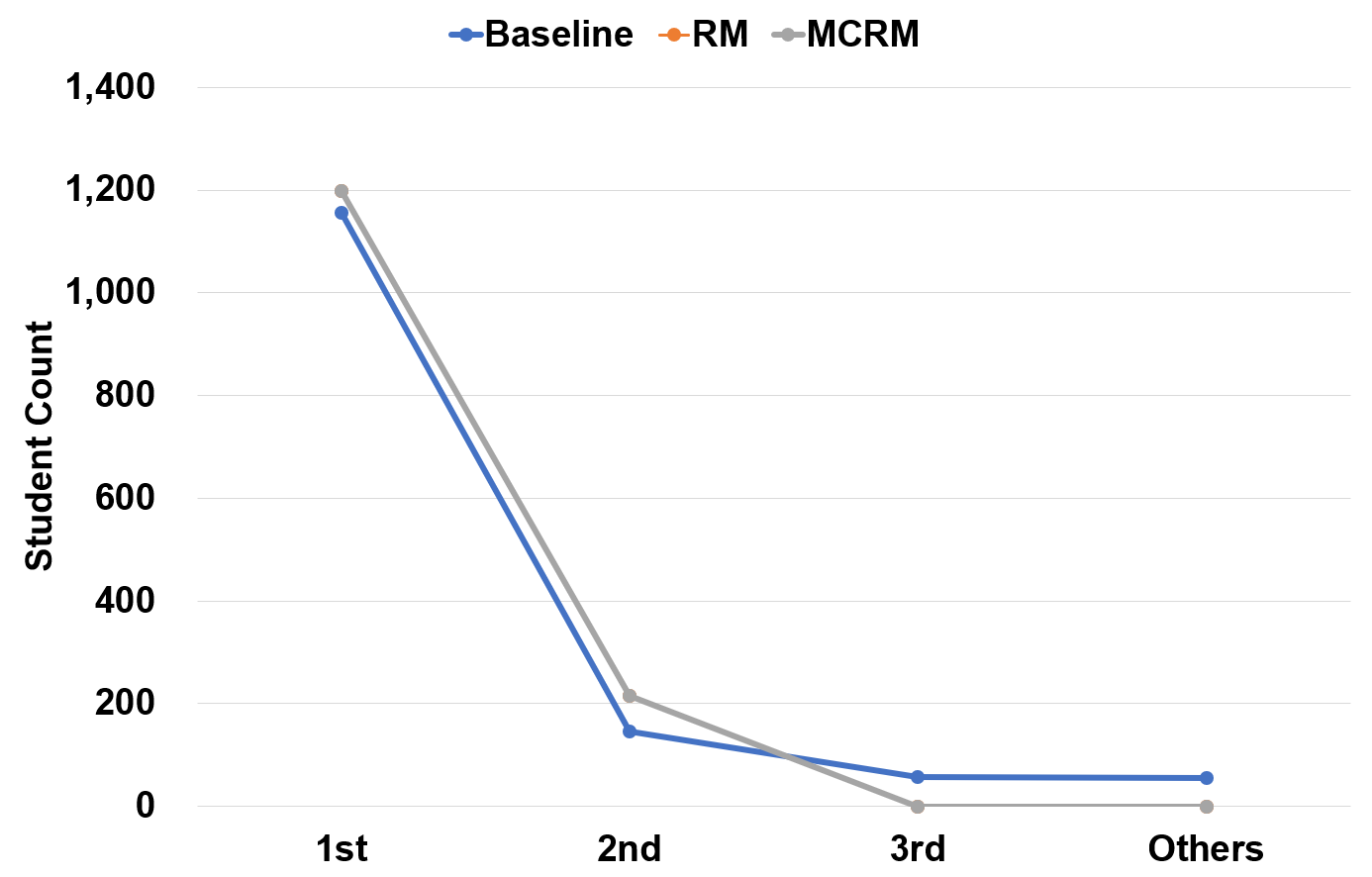}
    \caption{Profiles of Baseline, RM, and MCRM.}
    \label{fig:profile}
\end{figure}

Figure~\ref{fig:profile} shows the profiles of Baseline, RM, and MCRM.
Students who were assigned to another school than one of their three choices were counted as Others.
RM and MCRM, which achieved the same profile, so the lines overlap perfectly.
Since RM and MCRM assigned all students to their 1st or 2nd choice, there were no students assigned to their 3rd choice or Others.

\subsection{Commuting distance comparison}
\begin{table}[ht]
    \centering
    \begin{tabular}{c|c}
        \hline
        Algorithm   & Total commuting distance (km) \\
        \hline
        \hline
        Baseline    & 2624.11 \\
        RM          & 2679.23 \\
        MCRM        & 2514.53 \\
        \hline
    \end{tabular}
    \caption{Average of the total commuting distance.}
    \label{tab:comm_dist}
\end{table}

Table~\ref{tab:comm_dist} shows the total commuting distances for Baseline, RM, and MCRM.
MCRM minimized the total commuting distance, which decreased by 4.18\%, and 6.15\% compared to Baseline, and RM, respectively.
RM achieves the worst total commuting distance because, since RM only considers preferences, unlike Baseline or MCRM.









\bibliography{refs.bib}

\begin{thebibliography}{10}

\bibitem{berczi2022approximation}
Krist{\'o}f B{\'e}rczi, Tam{\'a}s Kir{\'a}ly, Yutaro Yamaguchi, and Yu~Yokoi.
\newblock Approximation by lexicographically maximal solutions in matching and
  matroid intersection problems.
\newblock {\em Theoretical Computer Science}, 910:48--53, 2022.

\bibitem{featherstone2020rank}
Clayton~R Featherstone.
\newblock Rank efficiency: Modeling a common policymaker objective.
\newblock {\em Unpublished paper, The Wharton School, University of
  Pennsylvania.[25, 28, 45]}, 2020.

\bibitem{gabow1989faster}
Harold~N Gabow and Robert~E Tarjan.
\newblock Faster scaling algorithms for network problems.
\newblock {\em SIAM Journal on Computing}, 18(5):1013--1036, 1989.

\bibitem{gai2007acyclic}
Anh-Tuan Gai, Dmitry Lebedev, Fabien Mathieu, Fabien De~Montgolfier, Julien
  Reynier, and Laurent Viennot.
\newblock Acyclic preference systems in p2p networks.
\newblock In {\em Euro-Par 2007 Parallel Processing: 13th International
  Euro-Par Conference, Rennes, France, August 28-31, 2007. Proceedings 13},
  pages 825--834. Springer, 2007.

\bibitem{huang2012weight}
Chien-Chung Huang and Telikepalli Kavitha.
\newblock Weight-maximal matchings.
\newblock {\em Proceedings of MATCH-UP}, 12:87--98, 2012.

\bibitem{huang2016fair}
Chien-Chung Huang, Telikepalli Kavitha, Kurt Mehlhorn, and Dimitrios Michail.
\newblock Fair matchings and related problems.
\newblock {\em Algorithmica}, 74(3):1184--1203, 2016.

\bibitem{irving2006rank}
Robert~W Irving, Telikepalli Kavitha, Kurt Mehlhorn, Dimitrios Michail, and
  Katarzyna~E Paluch.
\newblock Rank-maximal matchings.
\newblock {\em ACM Transactions on Algorithms (TALG)}, 2(4):602--610, 2006.

\bibitem{manlove2013algorithmics}
David Manlove.
\newblock {\em Algorithmics of matching under preferences}, volume~2.
\newblock World Scientific, 2013.

\bibitem{michail2007reducing}
Dimitrios Michail.
\newblock Reducing rank-maximal to maximum weight matching.
\newblock {\em Theoretical Computer Science}, 389(1-2):125--132, 2007.

\bibitem{ortega2023cost}
Josu{\'e} Ortega and Thilo Klein.
\newblock The cost of strategy-proofness in school choice.
\newblock {\em Games and Economic Behavior}, 141:515--528, 2023.

\end{thebibliography}

\end{document}


\maketitle

\begin{abstract}
프로필 기반 매칭 문제는 이분 그래프 $G = (A \cup B, E)$, $1 \le i \le r$에서 정의되는 공리 함수 $u_i: E \to \{ 0, 1, \dots, U_i \}$인 instance $(G, r, \langle u_1, \dots, u_r \rangle)$이 주어졌을 때, $r$-tuple로 표현되는 프로필을 최적화하는 매칭을 찾는 문제다.
공리 함수는 우선순위가 있어서, 최적의 매칭은 1번째 공리함수의 합을 최대화하고, 그 조건에서 2번째 공리함수의 합을 최대화하고, 등등을 만족하는 매칭이다.
프로필 기반 매칭 문제는 공리 함수에 따라 rank-maximal matching\cite{irving2006rank}, fair matching\cite{huang2016fair}, weight-maximal matching\cite{huang2012weight} 등을 표현할 수 있다.
이러한 문제들은 최대 가중치 매칭 문제로 환원이 가능하다고 알려져 있다.
그러나, 환원하는 방식은 매우 큰 가중치를 필요로 하여 복잡도가 매우 높은 것으로 알려져 있다.

본 논문에서는 프로필 기반 매칭 문제에서 최적의 매칭을 찾기 위한 가중치 함수의 조건을 제시한다.
mixed-radix number를 사용한 가중치 함수를 제시하고, 환원했을 때 문제의 복잡도가 $n = |V|, m = |E|$에서 $O(m\sqrt{n}(\log{n} + \sum_{i=1}^{r}\log{U_i}))$임을 증명한다.
또한, 가중치 함수의 조건을 기반으로 관련된 문제들의 새로운 가중치 하한을 제시하고, 일부 문제들의 복잡도를 개선하며, 조건을 만족하는 최대 가중치 매칭 문제의 복잡도가 $r^*$가 매칭에 포함된 rank 최대값일 때 $O(r^*m\sqrt{n})$임을 보인다.
끝으로, 학생들의 실제 선호도 데이터를 바탕으로 알고리즘의 효과성을 정량적으로 보인다.
\end{abstract}

\section{Introduction}
프로필 기반 매칭 문제는 이분 그래프 $G = (A \cup B, E)$와, 공리 함수의 개수 $r$, $1 \le i \le r$에서 최대값이 $U_1$인 non-negative 공리 함수 $u_i$가 입력으로 주어진다.
각 공리 함수는 도메인에 따라 선호도, 통학거리, latency와 같은 feature들을 정량화하는 함수다.
공리 함수는 우선 순위가 있으며, $u_1$이 가장 높은 우선 순위를 갖고, $u_2, \dots, u_r$순서로 우선순위가 낮아진다.
이러한 instance $(G, r, \langle u_1, \dots, u_r \rangle)$에서 최적의 매칭은 우선순위가 가장 높은 1번째 공리함수의 총합을 최대화하며, 그 조건에서 2번째 공리함수의 총합을 최대로 하며, 등등을 만족하는 매칭을 의미한다.


프로필 기반 최적의 매칭은 공리 함수에 따라 rank-maximal matching\cite{irving2006rank}, fair matching\cite{huang2016fair}, weight-maximal matching\cite{huang2012weight} 등을 표현할 수 있다.
이러한 문제들은 최대 가중치 매칭 문제로 환원이 가능함이 알려져 있다\cite{irving2006rank,huang2016fair,huang2012weight,manlove2013algorithmics}.
그런데 환원이 가능함을 직관적으로 보이기 위해 가파르게 증가하는 가중치를 사용하여 환원하는 알고리즘은 복잡도가 매우 높은 것으로 알려져 있다.
환원하는 방식의 정확한 복잡도를 얻기 위해 rank-maximal matching에 대해 환원하는 방식의 복잡도과 가중치의 하한에 대해서 연구가 되었다\cite{michail2007reducing,berczi2022approximation}

본 논문에서는 프로필 기반 매칭 문제에서 최적의 매칭을 찾기 위한 일반화된 가중치 함수의 조건을 제시한다.
프로필에 비례하도록 가중치를 설정하여 최대 가중치 매칭이 최적의 매칭이 됨을 증명한다.
이를 만족하는 mixed-radix number를 사용한 가중치 함수를 제시하고, 환원 알고리즘의 복잡도가 $n = |V|, m = |E|$에서 $O(m\sqrt{n}(\log{n} + \sum_{i=1}^{r}\log{U_i}))$임을 증명한다.

본 논문에서 제시한 가중치 함수의 조건을 기반으로 관련 문제들의 새로운 가중치 하한을 제시한다.
이러한 하한은 모든 관련 문제의 새로운 가중치 하한을 제시하며, 일부 문제들에 대해선 복잡도를 개선한다.
또한, 조건을 만족하는 최대 가중치 매칭의 복잡도를 $r^*$가 매칭에 포함된 rank 최대값일 때 $O(r^*m\sqrt{n})$으로 낮춘다.
끝으로, $\alpha$-city 학생들의 실제 선호도 데이터를 바탕으로 프로필 기반 매칭 알고리즘의 효과성을 정량적으로 보인다.

\begin{description}
\item[Contribution 1] 최대 가중치 매칭 문제로 환원하기 위한 가중치 조건을 제시함

본 논문에서는 문제를 단순화하기 위해 완전 이분 그래프에서 찾은 최적의 매칭이 원본 그래프에서 최적의 매칭임을 보인다.
완전 이분 그래프에서 프로필과 가중치 함수가 비례하도록 설계하면 최대 가중치 매칭이 최적의 매칭임을 증명한다.
이러한 가중치 조건은 관련 문제들에 대해 새로운 가중치 하한을 제시한다.
Rank-maximal matching\cite{berczi2022approximation}의 가중치 비율이 2보다 작을 수 있음을 예시를 통해 증명한다.
fair matching\cite{huang2016fair}과 weight-maximal matching\cite{huang2012weight}은 vertex 수 대신 (상수인) 공리의 최대값에 따라 결정되는 가중치 함수$w(e) = \sum_{i=1}^{r}{u_i(e) \prod_{j=i+1}^{r}{(2U_j+1)}}$를 제시한다.

\item[Contribution 2] 관련 문제들의 계산 복잡도를 개선함

본 논문에서는 프로필 기반 매칭을 최대 가중치 매칭 문제 환원하는 알고리즘을 통해 관련 문제들에 대해 더 나은 복잡도를 달성한다.
Fair matching과 weight-maximal은 복잡도가 매칭에 포함된 최대 순위가 $r^*$일 때 $O(r^*m\sqrt{n}\log{n})$으로 알려져 있다\cite{huang2016fair,huang2012weight,manlove2013algorithmics}.
반면 새로운 가중치 하한을 통해 환원하는 방식의 복잡도는 $O(m\sqrt{n}(\log{n} + r))$이다.
또한, 최대 가중치 매칭을 rank-maximal로 환원이 가능한지 효율적으로 알아내는 알고리즘을 제시한다.
조건을 만족하는 최대 가중치 매칭은 rank-maximal matching과 같은 복잡도인 $O(r^*m\sqrt{n})$으로 풀 수 있다.

\item[Contribution 3] 실제 데이터를 기반으로 profile-based matching들의 효과성을 보임

rank-maximal matching (RM)은 학교 배정 시 선호 만족 측면에서 배정 결과를 크게 개선하는 효과가 있다\cite{featherstone2020rank,ortega2023cost}.
본 논문에서는 $\alpha$-city의 실제 데이터를 기반으로, RM과, 비용으로 작용하는 통학거리를 최소화하는 minimum-cost rank-maximal matching (MCRM)을 정량적으로 평가한다.
실험 결과 RM과 MCRM은 모든 학생을 1지망 또는 2지망으로 배정한다.
MCRM은 가장 낮은 통학거리 총합을 달성하며, Baseline 대비 4.18\%, RM 대비 6.15\% 감소한다.
\end{description}

이어지는 섹션~\ref{sec:preliminaries}에서는 프로필 기반 매칭 문제와 증명에 필요한 개념들에 대해서 정의한다.
섹션~\ref{sec:reduction}에서는 최대 가중치 매칭 문제로 환원하기 위한 가중치 함수의 조건과, 이 조건을 만족하는 가중치 함수가 존재함을 보인다.
섹션~\ref{sec:complexity}에서는 환원하여 최적의 매칭을 찾는 알고리즘의 복잡도를 계산한다.
섹션~\ref{sec:related_problems}에서는 관련된 문제들에 대해서 가중치 하한 및 복잡도를 다룬다.
섹션~\ref{sec:experiment}에서는 실제 데이터를 사용한 실험 결과를 제시한다.
섹션~\ref{sec:conclusion}에서는 결론으로 마무리한다.

\section{Preliminaries}\label{sec:preliminaries}
프로필 기반 매칭 문제는 이분 그래프 $G = (A \cup B, E)$와, 각 edge에 연관된 $r$개의 공리 함수들$u_i$로 구성된 instance $(G, r, \langle u_1, \dots, u_r \rangle)$로부터 $r$-tuple로 표현되는 프로필을 최적화하는 매칭을 찾는 문제다.
\emph{공리 함수}는 $1 \le i \le r$에서 $u_i: E \to \{ 0, 1, \dots, U_i \}$으로 정의된다.
\emph{매칭}은 $E$의 부분 집합으로 어떤 edge도 서로에게 인접하지 않는다.
매칭 $M$에 대해 쌍 $(a, b) \in M$이면 $a$와 $b$는 서로에게 \emph{파트너}이며, $M(a) = b$, $M(b) = a$로 표기한다.
매칭 $M$의 \emph{프로필} $p(M)$은 $r$-tuple $\langle p_1(M), \dots, p_r(M) \rangle$로 정의되며, $1 \le i \le r$에서 $p_i(M) = \sum_{e \in M}{u_i(e)}$으로 정의된다.
매칭 $M, M'$에 대해 $i < j \le r$에서 $p_i(M) = p_i(M')$이고 $p_j(M) > p_j(M')$이면 $p(M) > p(M')$, $1 \le i \le r$에서 $p_i(M) = p_i(M')$ 이면 $p(M) = p(M')$으로 정의한다.
\emph{최적의 매칭}은 다른 모든 매칭보다 프로필이 크거나 같은 매칭으로 정의한다.

\begin{theorem}\label{theorem:perfect}
완전 이분 그래프 $G' = (A \cup B, A \times B)$와 $1 \le i \le r$에서 정의된 
공리함수들 $u_i'(e) = 0 \text{ for } e \notin E \text{ otherwise } u_i(e)$에서 생성된 최적의 매칭과 $E$의 교집합은, 이분 그래프 $G = (A \cup B, E)$와 공리함수들 $u_i$에 대해 최적의 매칭이다.
\end{theorem}

\begin{proof}
매칭 $M'$이 instance $(G', u_i', r)$의 최적의 매칭이며, 매칭 $M = M' \cap E$라고 하자.
매칭 $M$이 최적의 매칭이 아니라고 가정하면, instance $(G, u_i, r)$에는 매칭 $M$보다 프로필이 더 큰 매칭 $M^*$가 존재한다.

$\forall e \notin E,\ u_i'(e) = 0$이므로 다음 관계가 성립한다:
$$p(M') = p(M' \cap E) = p(M) < p(M^*)$$

$M^* \subset E \subset A \times B$이므로, instance $(G', u_i', r)$에는 매칭 $M'$보다 프로필이 큰 매칭 $M^*$가 존재한다.
그러나, 매칭 $M'$은 최적의 매칭이므로 $M'$보다 프로필이 큰 매칭은 존재하지 않는다.
매칭 $M$이 최적의 매칭이 아니라는 가정은 최적의 매칭보다 프로필이 큰 매칭이 존재한다는 모순을 만든다.
따라서, 완전 이분 그래프의 최적의 매칭과 $E$의 교집합은 원본 그래프의 최적의 매칭이다.
\end{proof}

\cref{theorem:perfect}에 의해 프로필 기반 최적의 매칭을 찾는 문제는 완전 이분 그래프에서 unacceptable pair의 공리가 0인 공리 함수를 가지고 최적의 매칭을 찾는 문제와 같아진다.
간결성을 위해 명시되지 않은 경우에는 완전 이분 그래프를 가정한다.

매칭 $M$을 \emph{개선하는 쌍} $(a, b)$는 $p(\{ (a, b) \}) > p(\{ (a, M(a)), (M(b), b) \})$을 만족하는 쌍으로 정의한다.
개선하는 쌍의 파트너를 서로 교체하여 개선하는 쌍이 포함된 새로운 매칭을 만드는 것을 매칭을 \emph{개선한다}고 한다.

\begin{lemma}
매칭을 개선하면 프로필이 증가한다.
\end{lemma}

\begin{proof}
쌍 $(a, b)$가 매칭 $M$을 개선하는 쌍이라 하자.
매칭 $M$과 개선된 매칭 $M'$은 다음 관계를 만족한다:
$$M \setminus \{(a, M(a)), (M(b), b)\} = M' \setminus \{(a, b), (M(b), M(a))\}$$
$$p(M) - p(\{(a, M(a)), (M(b), b)\}) = p(M') - p(\{(a, b), (M(b), M(a))\})$$

공리 함수는 non-negative function이므로, 개선하는 쌍의 정의에 의해 다음을 만족한다:
$$p(\{(a, M(a)), (M(b), b) \}) < p(\{(a, b)\}) \le p(\{(a, b), (M(b), M(a))\})$$
$$\therefore p(M) < p(M')$$

따라서, 개선된 매칭은 원본 매칭보다 프로필이 크다.
\end{proof}

\section{Reduction}\label{sec:reduction}
이 섹션에서는 프로필 기반 매칭을 maximum weight matching 문제로 환원하기 위한 가중치 함수의 새로운 하한을 제시한다.
매칭의 프로필이 클 수록 매칭의 가중치의 합이 커지도록 가중치 함수를 설계하면, 최대 가중치 매칭은 프로필이 가장 큰 매칭인 최적의 매칭이 된다.
매칭의 프로필은 개선할수록 증가하므로, 가중치 함수는 잠재적인 개선하는 쌍에게 더 큰 가중치를 부여해야 한다.

\begin{theorem}\label{theorem:weight}
$p(\{ (a, b) \}) > p(\{ (a, b'), (a', b) \}) \implies w((a, b)) > w((a, b')) + w((a', b)) \ge 0$을 만족하는 가중치 함수 $w$로 생성한 최대 가중치 매칭은 최적의 매칭이다.
\end{theorem}

\begin{proof}
가중치 함수 $w$으로 생성한 최대 가중치 매칭 $M$이 최적의 매칭이 아니라 가정하자.
매칭 $M$에는 개선하는 쌍 $(a, b)$가 있으며, 개선한 매칭을 $M'$이라 하자.

매칭 $M, M'$은 다음을 만족한다:
$$M \setminus \{(a, M(a)), (M(b), b)\} = M' \setminus \{(a, b), (M(b), M(a))\}$$
$$\sum\limits_{e \in M}{w(e)} - (w((a, M(a))) + w((M(b), b))) = \sum\limits_{e \in M'}{w(e)} - (w((a, b)) + w((M(b), M(a))))$$

쌍 $(a, b)$는 매칭 $M$을 개선하는 쌍이고, 가중치 함수 $w$는 non-negative function이므로 다음을 만족한다:
$$p(\{(a, M(a)), (M(b), b) \}) < p(\{(a, b)\})$$
$$w((a, M(a))) + w((M(b), b)) < w((a, b)) \le w((a, b)) + w((M(b), M(a)))$$
$$\therefore \sum\limits_{e \in M}{w(e)} < \sum\limits_{e \in M'}{w(e)}$$

최대 가중치 매칭 $M$보다 가중치 합이 큰 매칭은 존재하지 않는다.
그러나, 최대 가중치 매칭이 최적의 매칭이 아니라는 가정은 매칭 $M$보다 가중치 합이 큰 매칭 $M'$이 존재한다는 모순을 만든다.
따라서, 가중치 함수 $w$로 생성한 최대 가중치 매칭은 최적의 매칭이다.
\end{proof}

\begin{lemma}
Edge의 가중치를 base-$(2U_i+1)$의 mixed radix number로 다루는 가중치 함수 $w_{mixed}$는 $p(\{ (a, b) \}) > p(\{ (a, b'), (a', b) \}) \implies w_{mixed}((a, b)) > w_{mixed}((a, b')) + w_{mixed}((a', b)) \ge 0$을 만족한다.
\end{lemma}

\begin{proof}
가중치 함수 $w_{mixed}$를 공리 함수로 표현하면 다음과 같다:
\begin{align*}
\qquad\qquad\qquad\qquad
w_{mixed}(e) &= u_1(e)_{(2U_1+1)} u_2(e)_{(2U_2+1)} \dots u_r(e)_{(2U_r+1)} \\
&= \sum\limits_{i=1}^{r}{u_i(e) \prod\limits_{j=i+1}^{r}{(2U_j+1)}} \ge 0
\end{align*}

$p(\{ (a, b) \}) > p(\{ (a, b'), (a', b) \})$일 때 다음을 만족하는 정수 $x$가 존재한다:
$$u_i((a, b)) = u_i((a, b')) + u_i((a', b))\text{ for } i < x$$
$$u_x((a, b)) > u_x((a, b')) + u_x((a', b))$$

$x+1$ 이후 자리의 합은 $1_{(2U_{x}+1)} 0_{(2U_{x+1}+1)} \dots 0_{(2U_{r}+1)}$보다 작으므로, $w_{mixed}((a, b)) > w_{mixed}((a, b')) + w_{mixed}((a', b)) \ge 0$을 만족한다.
\end{proof}

\section{Complexity Analysis}\label{sec:complexity}
\begin{algorithm}[ht]
    \caption{An algorithm for finding a profile-based optimal matching.}\label{alg:reduction}
    \begin{algorithmic}[1]
        \Function{getProfileBasedOptimalMatching}{$V, E, r, \langle u_1, \dots, u_r \rangle, \langle U_1, \dots U_r \rangle$}
            \State $w \gets new\ Map(key\_type = Edge, value\_type = Integer)$ \label{code:weight}
            \ForEach{$e \in E$}
                \State $w[e] \gets \sum\limits_{i=1}^{r}{u_i(e) \prod\limits_{j=i+1}^{r}{(2U_j+1)}}$
            \EndFor \label{code:weight_done}
            \State $optimal\_matching \gets getMaximumWeightMatching(V, E, w)$ \label{code:mwm}
            \State \Return $optimal\_matching$
        \EndFunction
    \end{algorithmic}
\end{algorithm}

\cref{alg:reduction}은 최대 가중치 매칭 문제로 환원하여 최적의 매칭을 찾는 알고리즘을 보여주고 있다.
$n = |V|$, $m = |E|$라 하자.
Line~\ref{code:weight}-\ref{code:weight_done}는 공리를 가중치로 변환한다.
변환에는 edge의 공리들을 base-$(2 U_i+1)$인 mixed radix numeric system으로 간주하는 가중치 함수 $w_{mixed}$가 사용되었다.
모든 edge의 가중치를 계산하는데는 $O(rm)$이 필요하다.
Line~\ref{code:mwm}에서는 변환된 가중치를 사용하여 최대 가중치 매칭을 찾는다.
가중치의 최대값이 $N$일 때 복잡도는 $O(m\sqrt{n}\log{(n N)})$이다\cite{gabow1989faster}.
가중치의 최대값 $N < \prod_{i=1}^{r}{(2U_i+1)}$이므로, \cref{alg:reduction}의 복잡도는 $O(m\sqrt{n}(\log{n} + \sum_{i = 1}^{r}\log{U_i}))$이다.

\section{Related Problems}\label{sec:related_problems}
\subsection{Weight lower bound for rank-maximal matching}
서로 다른 rank의 edge는 가중치의 비율이 2보다 커야 rank-maximal matching을 찾을 수 있다고 알려져 있다 \cite{berczi2022approximation}.
하지만, 가중치 비율이 2보다 작아도 \cref{theorem:weight}을 만족하는 코너 케이스들이 있다.
예를 들어, 모든 edge들의 rank가 하나의 prefernce list로 결정되는 global preference system\cite{gai2007acyclic}에서 어떤 rank에 속한 모든 edge들은 하나의 공통된 vertex에 인접하는 경우를 가정하자.
가중치 함수 $w_{GRP}$를 rank가 $i$인 edge $e_i$에 대해 다음과 같이 정의하면, $r \ge 4$에서 $1 \le i \le r - 3$일 때 $\frac{w_{GRP}(e_{i})}{w_{GRP}(e_{i+1})}$는 2보다 작다.

\begin{align*}
\qquad\qquad\qquad\qquad\qquad
w_{GRP}(e_{i}) &= w_{GRP}(e_{i+1}) + w_{GRP}(e_{i+2}) + 1 \\
w_{GRP}(e_{r-1}) &= 2 \\
w_{GRP}(e_{r}) &= 1
\end{align*}

\subsection{Complexity of fair matching}
$r$까지 순위를 매기는 fair matching 문제\cite{huang2012weight,manlove2013algorithmics,huang2016fair}는 $r+1$개의 공리함수 $u_{fair, i}$는 $1\le j \le r$에서
$$u_{fair, 1}(e) = 1$$
$$
u_{fair, j+1}((a, b)) = \begin{cases}
2 & \text{if both $a$ and $b$ rank each other as rank $\le r - j + 1$}\\
1 & \text{if one of $a$ or $b$ ranks the other as rank $\le r - j + 1$} \\
0 & \text{otherwise}
\end{cases}
$$
으로 정의된다.
$a, b$가 edge $(a, b)$를 각각 $x, y$로 rank할 때, 가중치 함수로 $w_{fair}(e) = 4n^r - n^{x-1} - n^{y-1}$을 사용하면 maximum weight matching으로 환원이 가능하며, combinatorial technique을 사용하면 $O(r^*m\sqrt{n}\log{n})$의 복잡도를 갖는다고 알려져 있다\cite{huang2016fair}.
반면에, 본 연구에서 제시하는 가중치를 사용하면 공리 함수의 최대값은 1 또는 2이므로 $O(m\sqrt{n}(\log{n} + r))$의 복잡도로 fair matching을 찾을 수 있다.

\subsection{Complexity of weight-maximal matching}
Weight-maximal matching problem은 모든 공리 함수의 최대값이 $W$로 고정된 프로필 기반 매칭 문제다.
가중치 함수로 $w(e) = \sum_{j=1}^{r}{u_{WM, j}(e)n^{2(r-j)}}$을 사용하면 maximum weight matching으로 환원이 가능하며, combinatorial algorithm을 사용하면 $W$가 상수라는 가정 하에 $O(r^*m\sqrt{n}\log{n})$의 복잡도를 갖는다고 알려져 있다\cite{huang2012weight}.
반면에 본 연구에서 제시하는 가중치를 사용하면 $W$가 상수라는 가정 하에 $O(m\sqrt{n}(\log{n} + r))$의 복잡도로 weight-maximal matching을 찾을 수 있다.

\subsection{Complexity of maximum weight matching}
Rank-maximal matching은 maximum cardinality matching을 iterate하는 방식으로 풀기 때문에 환원보다 낮은 복잡도 $O(r^*m\sqrt{n})$를 갖는다\cite{irving2006rank}.
조건을 만족하는 최대 가중치 문제를 효율적으로 rank-maximal로 환원하면, 복잡도를 크게 개선할 수 있을 것이다.
\cref{alg:sort_sweep}은 최대 가중치 매칭 문제가 rank-maximal matching의 가중치의 조건을 만족하는지는 확인하는 알고리즘이다.

\begin{algorithm}[ht]
    \caption{An algorithm to verify the rank-maximal condition.}\label{alg:sort_sweep}
    \begin{algorithmic}[1]
        \Function{isRankMaximal}{$V, E, w$}
            \State $weight\_list \gets new\ List(type = Integer)$ \label{code:sort}
            \ForEach{$e \in E$}
                \State $weight\_list.append(w(e))$
            \EndFor
            \State $sorted\_list \gets sort(weight\_list, increasing = True)$ \label{code:sort_done}
            \State $\langle w_i, w_j, w_k \rangle \gets \langle sorted\_list[0], 0, 0 \rangle$ \label{code:sweep}
            \For{$i \gets 1 \text{ to } |E|$}
                \State $\langle w_i, w_j, w_k \rangle \gets \langle sorted\_list[i], w_i, w_j \rangle$
                \If{$w_i \le w_j + w_k$}
                    \State \Return $False$
                \EndIf
            \EndFor \label{code:sweep_done}
            \State \Return $True$
        \EndFunction
    \end{algorithmic}
\end{algorithm}

Line~\ref{code:sort}-\ref{code:sort_done}에서 가중치를 정렬하는 복잡도는 $m \le (\frac{n}{2})^2$이므로 $O(m \log{n})$이다.
\cref{code:sweep,code:sweep_done}에서 정렬된 가중치를 sweep하며 rank-maximal의 조건을 만족하는지 확인하는 복잡도는 $O(m)$이다.
따라서, \cref{alg:sort_sweep}의 전체 복잡도는 $O(m \log{n})$이다.
만약 최대 가중치 매칭 문제의 가중치가 rank-maximal matching의 조건을 만족한다면, 가중치가 가장 큰 edge들을 rank 1, 다음으로 큰 edge들을 rank 2, ... 가장 가중치가 작은 edge들을 rank $r$로 설정한 rank-maximal matching으로 환원하면 $O(r^*m\sqrt{n})$로 풀 수 있다.

\section{Experimental Results}\label{sec:experiment}
In this section, the performance of matching algorithms is evaluated with real data on students' preferences, commuting distances, and schools' quotas which was used for actual school assignment in $\alpha$-city in 2023.
There are nine schools in the city, and each school has separate quotas for males and females.
Table~\ref{tab:quota} shows the number of male and female students in each school.

\begin{table}[ht]
    \centering
    \begin{tabular}{|c|c|c|c|}
        \hline
        School  & Male  & Female    & Total \\
        \hline
        \hline
        $h_1$   & 154   & 0         & 154   \\
        $h_2$   & 161   & 0         & 161   \\
        $h_3$   & 172   & 0         & 172   \\
        $h_4$   & 0     & 186       & 186   \\
        $h_5$   & 0     & 170       & 170   \\
        $h_6$   & 0     & 172       & 172   \\
        $h_7$   & 80    & 73        & 153   \\
        $h_8$   & 77    & 46        & 123   \\
        $h_9$   & 71    & 52        & 123   \\
        \hline
        Total   & 715   & 699       & 1414  \\
        \hline
    \end{tabular}
    \caption{School quotas in $\alpha$-city in 2023.}
    \label{tab:quota}
\end{table}

Students applied three schools.
Students who are not assigned to a school to which they applied due to insufficient seats may be assigned to another school to which they did not apply.
In other words, schools not included in the choices are considered as fourth choices, which forms complete preferences with ties.
Commuting distance is calculated by using the walking distance\footnote{The average of the recommended walking distance on Naver Map and the shortest walking distance on Kakao Map is measured in 0.01km units} from the student's residence to the school.

All experiments were performed using Python 3.10 on a system with AMD Ryzen 7 7735HS and 16GB of DDR4 RAM.
The results of the simulations were averaged using 10 randomly selected seeds to shuffle the order of iterating students.
DA that requires tie-breaking was simulated after tie-breaking with 10 separately selected random seeds, so it averages 100 results.

\subsection{Baseline: Actual assignment result in $\alpha$-city in 2023}
$\alpha$-city assigned students to schools in two stages.
In the first stage, 80\% of the seats in each school were allocated based on the students' preferences using a greedy algorithm.
In the next stage, the schools were ranked in six ranks from the nearest school to the furthest school for each student, and the remaining 20\% of seats were assigned using a greedy algorithm using the rank determined by distance.

\subsection{RM: Rank-maximal matching}
The rank-maximal matching problem can only consider preferences, so commuting distance is not taken into account.
The weight function $w_{RM}(e) = 2^{4 - rank(e) + 1} - 1$ is used for RM

\subsection{MCRM: Minimum-cost rank-maximal matching}
MCRM은 RMs 중 통학거리가 최소인 매칭을 의미한다.
통학거리 함수 $d: E \to \{0, 1, \dots, D\}$에 대해 가중치 함수는 $w_{MCRM}(e) = (D+1)2^{4 - rank(e)} - D - d(e)$을 사용했다.

\subsection{Profile comparison}
\begin{figure}[ht]
    \centering
    \includegraphics[width=0.95\linewidth]{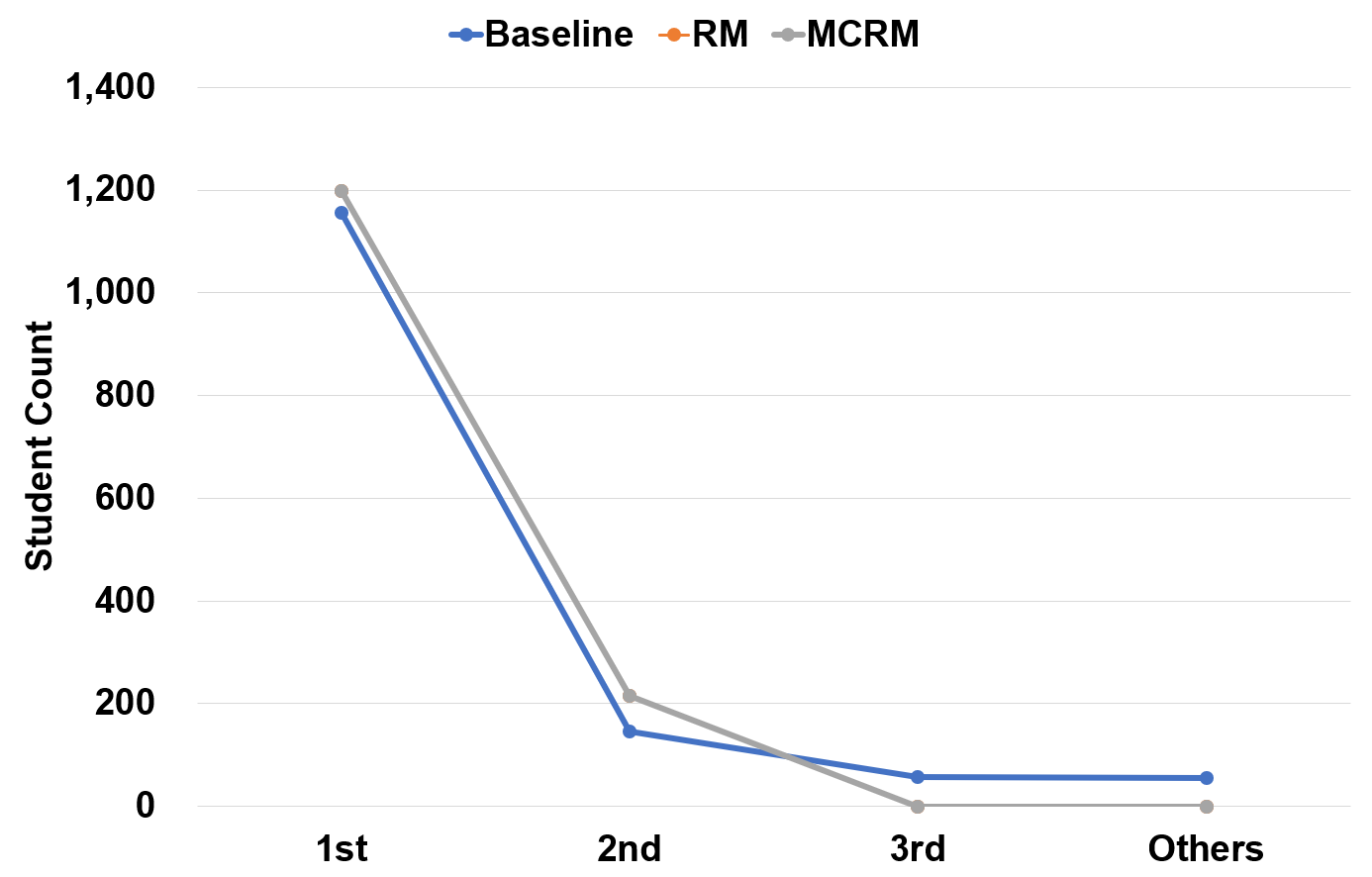}
    \caption{Profiles of Baseline, RM, and MCRM.}
    \label{fig:profile}
\end{figure}

\cref{fig:profile}은 Baseline, RM, MCRM의 profile을 보여주고 있다.
자신이 선택한 세 학교 중 하나가 아닌 다른 학교에 배정된 학생은 기타로 계산된다.
RM과 MCRM은 profile이 동일하여 선이 완벽히 겹쳐져 있다.
RM과 MCRM은 1지망 또는 2지망에 모든 학생들을 배정하여 Baseline보다 높은 profile을 달성했다.

\subsection{Commuting distance comparison}
\begin{table}[ht]
    \centering
    \begin{tabular}{c|c}
        \hline
        Algorithm   & Total commuting distance (km) \\
        \hline
        \hline
        Baseline    & 2624.11 \\
        RM          & 2679.23 \\
        MCRM        & 2514.53 \\
        \hline
    \end{tabular}
    \caption{Average of the total commuting distance.}
    \label{tab:comm_dist}
\end{table}

\cref{tab:comm_dist}는 Baseline, RM, MCRM의 통학거리의 총합을 보여주고 있다.
MCRM은 Baseline보다 4.18\%, RM보다 6.15\% 감소했다.
RM은 Baseline보다 2.10\% 증가했는데, 통학거리를 고려하는 Baseline과 달리 RM은 선호도만 고려하기 때문이다.

\section{Conclusion}\label{sec:conclusion}
본 논문에서는 프로필 기반 매칭 문제를 최대 가중치 매칭 문제로 환원하는 방법을 제시함

완전 이분 그래프로 변환해도 프로필이 변하지 않도록 공리 함수를 설정

개선하는 쌍이라는 개념을 통해 가중치 함수의 조건 제시

mixed radix number를 사용하여 가중치 함수를 제시

조건을 만족하는 가중치 함수는 RM의 새로운 가중치 하한

FM, WM의 복잡도를 낮출 수 있음을 증명

조건을 만족하는 최대 가중치 매칭 문제는 복잡도를 RM과 같도록 낮출 수 있음을 증명

실제 데이터를 기반으로 프로필 기반 매칭의 효과성을 보임

\bibliography{refs.bib}